# KURT GÖDEL AND HIS UNIVERSE[*]

Ivan Todorov

Institute for Nuclear Research and Nuclear Energy
Bulgarian Academy of Sciences
e-mail: todorov@inrne.bas.bg

A portrait of Kurt Gödel with emphasis on his work on *relativity theory and idealistic philosophy.*

Experts in mathematical logic, including Solomon Feferman, editor of Gödel's collected work [G], go out of their way, to point out that the significance of Gödel's incompleteness theorem in pure mathematics (outside logic), not to speak of other fields, like natural philosophy, is been exaggerated[1]. In his perceptive biographical memoir, the logician Georg Kreisel writes, for instance ([K80] p. 149): "Despite sensational presentations by crackpots, philosophers and journalists, Gödel's results have not revolutionized the silent majority's conception of mathematics, let alone its practice." (I wonder whether Kreisel would extend his crackpot's label to authors like Penrose, [P89, 94], or his opponent Nelson, [N], who keep discussing the philosophical implications of Gödel's work.) The truth is that Gödelquantum gravity himself thought and worked (if not published) most of his life on the philosophy of logic, mathematics, physics, theology. There is a unity in his life-work. His study of the problem of time in general relativity belongs to a period he is immersed into idealistic philosophy – from Parmenides (6$^{th}$-5$^{th}$ century BC) through Leibniz (1646-1716) to Kant (1724-1804) and Husserl (1859-1938).

Sect. 1 provides a glimpse into Gödel's early life and the work in Vienna that brought him world fame. His work on rotating universes in general relativity, which belongs to his *philosophical* (Princeton) *period*, is surveyed in Sect. 2. Sect. 3 tells the story of Gödel's friendship with Einstein; it offers an impression of his last years and ends with a few words about his *Nachlaß*.

## 1. "Der Herr Warum" enters "der Wiener Kreis"

*Die Welt ist vernünftig.* (The world is rational.)
Kurt Gödel (undated, [D97], p. 1)

Born (April 1906 in Brünn/Brno, Moravia) in the Austrian-Hungarian empire Gödel is treated, after the defeat in 1918 of the Central Powers, as a citizen of Czechoslovakia. In spite of living in a Catholic state and environment, Kurt, as well as his elder brother Rudolf (1902-94), is baptised in a Lutheran congregation, following the religion of his German mother (rather than that of his Austrian father). He is an exceptionally inquisitive child: by the age of four his parents call him "der Herr Warum" (Mr. Why). According to psychologists children ask questions that adults regard as not having answers until they get accustomed with the idea of chance. Gödel never stops asking such questions. He does not accept the notion of fortuitous events. The science to which he devotes his life confirms it; as he will later write, "In the world of mathematics everything is well posed and in perfect order. Should not the same be expected for the world of reality, contrary to appearances?" ([D97], p. 2) Asking unanswerable questions, seeking a rational explanation for everything contributes to his social isolation: it is considered *irrational*.

He is remembered as a generally happy but rather timid child, unusually troubled when his mother leaves the house. Having suffered from a rheumatic fever and reading medical books (at the age

---

of eight), concerns for his health begin to take up more and more of his daily life. A fellow logician, also prone to fears about health, finds it natural that his first romantic interest (cut short by his parents) is the daughter of family friends, ten years older than him ([K80], pp. 152-153).

In Brno's *Realgymnasium* he excels in theology and the languages – Latin, French, English (besides the native German; he neither studies nor speaks Czech, however, viewing himself, after 1918, as an Austrian exile; his interest in languages – at least in their formal aspects – continues beyond the school years). The only less than highest mark, he once gets, is in ... mathematics ([D97] p. 15).

In 1924 Kurt joins his brother (who studies medicine) in Vienna as a physics student. The lectures of the number theorist Furtwängler[2], *the most wonderful he ever heard*, makes him switch to mathematics. His professor (later thesis-advisor) Hahn[3] has been instrumental in bringing the (physicist and) philosopher Schlick[4] to Vienna. Each Thursday they get together with a small group of followers in an old Viennese coffee-house. Such are the beginnings of the celebrated Vienna Circle (*Der Wiener Kreis,* title of a 1929 manifesto). Hahn, who also invites the twenty-year-old Gödel to the meetings, directs the group's attention (then focussed on Wittgenstein's *Tractatus*) towards (mathematical) logic.

These are, perhaps, the happiest years of Gödel's life: respected by professors and colleagues, stimulated by discussions in the Circle, he starts his work on the doctoral dissertation, and, no less important, he meets Adele Porkert (1899-1981), his future wife. His parents, especially his father, once again object: she is a dancer, six years older, divorced... Kurt, now twenty-one, does not break with Adele, but he avoids conflicts and they only marry in 1938, nine years after his father's death (when his mother is not in Vienna either). Forty years later Gödel's junior colleague will write: "I visited them quite often in the fifties and the sixties. It was a revelation to me to see him relax in her company. She had little formal education but a real flair for the *mot juste*, which her somewhat critical mother-in-law eventually noticed too, and a knack for amusing twists on a familiar ploy: to invent far fetched grounds for jealousy. On one occasion she painted IAS /the Institute for Advanced Study in Princeton/, which she usually called *Alterversorgungsheim* (home for old-age pensioners), as teeming with pretty girl students who queued up at the office doors of permanent professors. Gödel was very much at ease with her style." ([K80] pp. 154-155).

Gödel, like his mentor Hahn, is an admirer of Leibniz, the first philosopher that has gone beyond Aristotle's (384-322) syllogisms (by contrast not only to his predecessors but also to such later figures as Kant, who writes that the logic "neither had to retrace nor been able to advance" a single step since Aristotle). Leibniz's idea of denoting primitive concepts by prime numbers and combinations of them by composite integers, is later exploited by Gödel. The interest in mathematical logic, as witnessed by the 1900 Hilbert's problems and by his program outlined at the 1928 Bologna Congress of Mathematicians, is triggered by the theory of (infinite) sets and associated *transfinite numbers*.

This is not the place to tell the story of "conquering infinity" in mathematics and logic. Just a few words about, arguably, the most fascinating predecessor of Gödel: "Cantor's grand meta-narrative, Set Theory, created by him almost single-handedly, resembles a piece of fine art more than a scientific theory" [M02]. His 1883 monograph on the foundations of set theory, subtitled *Ein matthematisch-philosophischer Versuch in der Lehre des Unendlichen* (A Mathematical-Philosophical Attempt in the Study of the Infinite), his first great work on the topic, has mathematical and philosophical sections inextricably connected ([D90] Ch. 6). Cantor states but cannot prove his continuum hypothesis. He believes that infinite sets exist as fully realised entities in the mind of God ([Y02] pp. 34-36). In 1884 he has a mental breakdown. He blames the strain to his mathematical work and becomes more passionate in his studies of the Scriptures. Returning to mathematics he obtains new important results: he introduces his diagonalization procedure and proves very generally that the set of all subsets of a set is larger than the set itself. In 1895-97 he publishes his last major mathematical work. But he also becomes aware of paradoxes[5] in set theory. The ghost of his disease comes back. Life becomes a balancing act. Repeatedly confined to Halle *Nervenklinik*, Cantor, 73, dies there in 1918, during the last (cold and hungry) war winter (*to escape,* as he writes in a poem, *the world I am in*).

---

2  Philipp Furtwängler (1869-1940), one of the founders of class field theory, lectured from his wheel chair without notes, while a scribe wrote on the board; ... an exceptionally fine head reminding of his cousin, the conductor, [K80] p. 153.
3  Hans Hahn (1879-1934), student of Mach in Vienna, best known for the Hahn-Banach theorem in functional analysis.
4  Moritz Schlick, born in Berlin, 1882, student of Max Planck (1858-1947), occupies (since 1922) the chair in Philosophy and Inductive Sciences in Vienna University, once held by Ludwig Boltzmann (1844-1906) and by Ernst Mach (1838-1916); influenced by Wittgenstein (1889-1951) and a leader of *der Kreis*, he is shot dead by a deranged student in 1936.

Gödel's glamorous achievements, the *completeness theorem* (his dissertation, 1929) and the *incompleteness theorem* (1931, his *Habilitationsschrift*) have been recounted - within their historical context, in an increasing level of detail and sophistication – e.g. in [Y02] [F86] [F06] [ML87] [K80] [D97] [W87] as well as in his collected work with commentary, [G]. It is for this theorem that Gödel is referred to as *the discoverer of the most significant mathematical truth in the century* (on the occasion of his Honorary Doctor degree from Harvard University in 1952 – [W87] p. XXIII). To quote an assessment by a peer ([K80] p. 150): the main ingredient in Gödel's incompleteness result is the *"philosophical distinction between arithmetic truth and derivability"*; he discerns a general pattern: *"by attention to philosophical notions and issues, adding possibly a touch of precision, one arrives painlessly at appropriate concepts, correct conjectures, and generally easy proofs"*. We shall see that Gödel follows a similar pattern in his work on the problem of time in general relativity.

Many, following von Neumann (1903-1957), call Gödel the greatest logician since Aristotle (384-322). Nelson's words, "The logic of Aristotle – the greatest logician before Gödel – is inadequate for mathematics." (*Mathematics and Faith*, [N]), are more precise: Gödel rather than Aristotle is the measure of greatness. Kreisel ([K80] p. 219) adds: "if Gödel's work is to be compared with that of one of the ancients, *Archimedes* (287-212) *is a better choice than Aristotle. Archimedes did not invent mechanics, as Gödel did not invent logic. But they both changed their subjects profoundly by work with almost unsurpassable ratio of interest of the result to effort.*"

The great incompleteness result, in spite of its technical simplicity, takes some time before being appreciated by the experts in the field. Zermelo (1871-1953), to whom we owe the modern axioms of set theory (then sixty, having suffered a nervous breakdown), challenges it in print (and never understands it – in spite of Gödel's patient effort to explain it to him, [D97], pp. 75-77). Even Hilbert (1862-1943), whose program triggers Gödel's work, has a negative initial reaction, albeit a few years later he works on it (together with his younger collaborator, Paul Bernays, who has consulted Gödel in person). (Only von Neumann, who has been thinking himself about the problem, grasps the result and its significance immediately – and helps securing Gödel a place at IAS.) In any case, by 1935, the incompleteness theorem has changed the outlook of mathematical logic, and another strange young man, Alan Turing (1912-1954), is making the next important step. We shall see that theoretical physicists will take longer to absorb Gödel's contribution to general relativity.

## 2. Exodus. Philosophy and physics

*No reason can be given why an objective lapse of time should be assumed at all.*
Kurt Gödel [G49b]

Men of science have difficulty in making decisions. The vacillations of Hermann Weyl (1885-1955) in the course of emigrating from Göttingen to Princeton in 1933 are documented in Chapter 6 of [B06] (pp. 134-137, 145-148). Gödel only makes up his mind to leave Vienna in the fall of 1939. He is "helped" in that by an assault of young Nazi rowdies in the vicinity of the University; the youths (having mistaken him, perhaps, for a Jew) knock off his glasses before Adele manages to drive them away with her umbrella ([D97] p. 147). As World War II is going on, the couple has to take a long rail route via Siberia and a liner crossing the Pacific in early 1940. The thirty-four-year-old Gödel will stay at the Institute for Advanced Study (IAS) in Princeton for the remaining 38 years of his life. One of the first persons who meets him there, his countryman Morgenstern[6], writes in his diary: "Gödel has come from Vienna. ... In his mix of profundity and otherworldliness he is very droll. ... When questioned about Vienna, he replied 'The coffee is wretched'" ([D97] p. 153).

---

5  By 1961 Gödel will be entitled to say: "... the antinomies of set theory, contradictions, ... whose significance was exaggerated by skeptics and empiricists ... have been resolved in a manner that is completely satisfactory and, for everyone who understands the theory, nearly obvious" (quoted after Yandell, [Y02] p. 51).

In the first two years in America Gödel continues his foundational work on set theory – trying to prove the independence of Cantor's continuum hypothesis. Exhausted and only half successful (the proof is completed by Paul Cohen in 1963) by the fall of 1942 he turns to philosophy. It is our good fortune that he is then invited[7] to contribute to a volume dedicated to Russell in the prestigious *Library of Living Philosophers*. Near the end of the article (completed in 1943) he expresses frustration that Leibniz's vision of logic "like a jewel that can throw light in many different directions" ([W87] p. 261) is not yet coming true: *Many symptoms show only too clearly, however, that the primitive concepts need further elucidation. It seems reasonable to suspect that it is this incomplete understanding of the foundations which is responsible for the fact that Mathematical Logic has remained so far behind the high expectations of Peano and others who (in accordance with Leibniz's claims) had hoped that it would facilitate theoretical mathematics to the same extent as the decimal system of numbers has facilitated numerical computations. For how can one expect to solve mathematical problems by mere analysis of the concepts occurring, if our analysis so far does not even suffice to set up the axioms?*

In 1946 Gödel is again invited to contribute to the Library of Living Philosophers, now to a volume dedicated to Einstein. He promises to write a short essay entitled "The theory of relativity and Kant". It appears that he quickly becomes absorbed with the question since a month later he writes to his mother that he is so deeply involved in his work that he finds it hard to summon the concentration for writing letters. It is clear from the outset that he is led by his philosophical interest in the concept of time. He is attracted by Kant's idea[8] that the notion of a time interval is subjective and he sets himself to find its confirmation in the general theory of relativity. Working with characteristic intensity, he constructs, by the summer of 1947, a solution describing a rotating Universe which does not admit a global notion of time and simultaneity. Still unhappy with the ready manuscript (resisting the editor's attempts to extract it from him), he continues to work throughout the next year. He apologizes to his mother (in May 1948) for a two month delay of his reply to her letter: a problem has driven everything else out of his mind. Even when he listens to the radio he is doing it "with only half an ear". This is the period when Gödel discovers that his solution admits closed time-like lines, that is, there exists, at least theoretically, a possibility to revisit one's own past. Taking his solution seriously, he addresses (in a manuscript published posthumously in [GIII]) the implied paradox: can one, visiting his past, try to alter it? Such inconsistency, he writes, *presupposes not only the practical feasibility of* such a *trip* (for which *velocities very close to the speed of light would be necessary*) *but also certain decisions on the part of the traveller, whose possibility one concludes only from a vague conviction of the freedom of the will* ([D97] p. 183). (Adele jokes that Kurt is taking the idea of communicating with one's past so close to heart that he keeps reading books about ghosts, [K80] p. 155.)

The volume is not ready for Einstein's seventieth birthday (March 1949): Gödel only hands his essay (in person) at the gala celebration. He publishes three (short) articles on the subject: [G49a,b] and [G50/52]. The style is hard to imitate. Here is a quotation from the philosophical essay, [G49b]:

---

6 The economist Oscar Morgenstern (1902-1977) is one of the few friends of Gödel in Princeton. Concerning his diary – see J. W. Dawson, Jr., In quest of Kurt Gödel: Reflections of a biographer, *Notices of the AMS* **53**:4 (2006) 444-447.

7 *All* Gödel's papers of his American period are invited. That is even true for some of his unpublished manuscripts, like the Gibbs Lecture: *Some basic theorems on the foundations of mathematics and their philosophical implications*, and his Carnap's paper: *Is mathematics syntax of language?,* different version of which have appeared in 1995 ([G95], [GIII]).

8 Gödel is not the first great mathematician influenced by Kant's concept of time. W. R. Hamilton (1805-1865) is entitling a major work of his as "... Essay on algebra as the science of pure time" (referring to Kant) – see [H80], Chapter 17.

*Relativity theory gave new and surprising insights into the nature of time, of that mysterious and seemingly self-contradictory being which, on the other hand, seems to form the basis of the world's and our own existence. The very starting point of special relativity theory consists of a new and very astonishing property of time, namely the relativity of simultaneity, which to a large extent implies that of succession.* It appears to deprive *the lapse of time* of its objective meaning. The existence of matter however, Gödel continues, distinguishes the observers who *follow ... the mean motion of matter*. In earlier cosmological solutions *the local times of these observers fit together into one world time.* In his new solution (given in [G49a]) there is no such *world time*. Moreover, *by making a round trip on a rocket ship in a sufficiently wide curve, it is possible in these worlds to travel into any region of the past, present, and future, exactly as it is possible in other worlds to travel to distant parts of space*. Hence, *the experienced lapse of time can exist without an objective laps of time*. For Gödel this supports *the view of those philosophers who, like Parmenides[9], Kant and the modern idealists, deny the objectivity of change as an illusion or an appearance due to our special mode of perception* (cf. the discussion in [W87] pp. 182-185). In his published remarks in the volume containing the Gödel's essay Einstein acknowledges that the possibility of closed time-like lines "disturbed me at the time of the building up of the general theory of relativity." Having failed to clarify the question himself, he hails Gödel's discovery as "an important contribution". But he does not fully trust the consequences of his own theory and adds: "It will be interesting to weigh whether these are not to be excluded on physical grounds." Gödel, on the other hand, states (concluding [G49b]): *The mere compatibility with the laws of nature of worlds in which there is no distinguished time, and therefore no objective lapse of time exists, throws some light on the meaning of time also in those worlds in which an absolute time can be defined.* In other words, as *the laws of nature*, incorporated in general relativity, allow for the solution under consideration, no *ad hoc* "physical grounds" can exclude it.

It is surprising (also to Gödel) that these basic questions have not been settled thirty years after the creation of general relativity, and that an "outsider" has to advance them. In fact, solutions involving rigidly rotating perfect fluids have appeared earlier in the literature but neither the existence of closed time-like curves has been noticed nor the problem of the existence of a global time is been discussed (for a review - see [I02] where the early papers of Lanczos (1924; English translation, 1997), Lewis (1932) and van Stockum (1937) are cited). It also takes time, as we shall see, to understand and absorb Gödel's discoveries.

On 7 May 1949 Gödel lectures about his work in IAS (to Einstein, Oppenheimer, Veblen, Chandrasekhar, Chern, among others). He speaks for an hour and a half - "in good form" but over the heads of most of his audience ([D97] p. 184). Many are astonished by his knowledge of physics. Nonetheless, the correctness of his cosmological results (like that of the incompleteness theorem) is challenged in print, 12 years later, by one of the most knowledgeable persons in the audience[10]. It takes another eight years, during which the possibility of time travel is "treated as doubtful in the philosophical literature", before someone (H. Stein, who has to appeal to the support of Gödel to get his paper published) points out that the criticism is based on a misunderstanding (Chandra mistakenly imputes to Gödel the statement that the closed timelike curve is a geodesics). Gödels' talk at the 1950 International Congress of Mathematicians, [G50/52], is met with ovation, but his work begins to be understood and appreciated long after his death, nearly half a century after it has first been reported[11].

The absence of an objective notion of "time of the Universe" in general relativity being settled in principle, Gödel is not indifferent to what is the solution chosen by Nature. To take the observed red shift into account he sketches in [G50/52] the possibility of an expanding rotating solution which (albeit it does not involve closed time-like lines) still admits no global absolute time (and violates "Mach's principle"). He points out that *a directly observable necessary and sufficient condition for the rotation of an expanding spatially homogeneous and finite universe* is that *for sufficiently great distances there must be more galaxies in one half of the sky than in the other half.* Gödel repeatedly asks his IAS colleague, Freeman Dyson[12], about the observational evidence for rotation, but he does not seem to trust the astrophysicists alone: two bounded notebooks are found in his *Nachlaß* in which he records himself angular orientations of galaxies ([D97] p. 182).

---

9 *The Greek were not addicted to moderation*, writes Bertrand Russell, *Heraclitus maintained that <u>everything</u> changes; Parmenides retorted that <u>nothing</u> changes* ([R], Chapter V, p. 48).
10 S. Chandrasekhar (1910-1995, Nobel Prize in astrophysics, 1983) and J. P. Wright publish their criticism in 1961. As late as 1978, 29 years after Gödel's lecture on the subject, a note from the director of IAS on who should speak about what at Gödel's funeral says "relativity not worth a talk" (handwritten note reproduced in [SDM], p.151).
11 Closed timelike curves (often abbreviated as CTC) became a respectable topic, starting (apparently) with the discussion about their relation to cosmic strings [G91] [DtH92] and continuing to attract attention to date (see also the book [Y05]).

The problem of defining time in general relativity attracts since much attention. An interesting suggestion, [R93] [CR94], consists in introducing a *thermodynamic time* attached to a state in thermal equilibrium (in which, substantially, *nothing changes*). Still, it does not seem, that the last word on this problem and, more generally, on the significance of Gödel's work has been pronounced.

### 3. Gödel and Einstein. Epilogue

> Einstein told me that his own work no longer meant much, that he came to the Institute merely *um das Privileg zu haben, mit Gödel zu Fuss nach Hause gehen zu dürfen* (to have the privilege to be able to walk home with Gödel)  O. Morgenstern, Letter to Bruno Kreisky, 1965 ([W87] p. 31)

Princeton society proved unreceptive (if not hostile) towards Adele and she appears to have led a lonely life there. (She is overheard to beg Kurt to accept an offer to move to Harvard as they are "so much more friendly" - [W87] p. 100.) To Gödel it is different. *The Institute harbours another philosophically minded German-speaking refugee, past his scientific prime: Einstein. The two most uncommon men find a friendship that gives both of them solace* ([Y02] p. 55). Here are some witnesses:

"The one man who was, during the last years, certainly by far Einstein's best friend, and in some ways strangely resembled him, was Kurt Gödel, the great logician. They were very different in almost every way – Einstein  gregarious, happy, full of laughter and common sense, and Gödel extremely solemn, very serious, quite solitary, and distrustful of common sense as a means of arriving at the truth. But they shared a fundamental quality: both went directly and wholeheartedly to the questions at the very centre of things" - writes Ernst Gabor Straus (1922-1986), Einstein's assistant in Princeton, 1944-48, later professor in Los Angeles (quoted in [W87] p. 115). Further (Straus again): "Einstein ... felt that he should not become a mathematician because the wealth of interesting and attractive problems was so great that you could get lost in it. In physics he could see what the important problems were and could, by strength of character and stubbornness, pursue them. But he told me once '*Now that I've met Gödel, I know that the same thing does exist in mathematics*' ([W87] pp. 31-32). In spite of being close to each other it is not clear whether Gödel discussed his work on general relativity with Einstein prior to its completion. As witnessed by Straus, *Gödel was totally solitary and would never talk to anybody while working*.

Morgenstern's story of how Gödel gets his American citizenship has been told more than once (one sees, among other things, how his friends are trying to protect him – as one would do it with a child).

Gödel is to take the routine citizenship examination, and he prepares for it very seriously studying the United States Constitution. On the eve, he tells Morgenstern that he had discovered a logical-legal possibility of transforming the United States into a dictatorship. Morgenstern sees that such a hypothetical possibility (and the complex chain of reasoning leading to it) is not suitable for discussing in an interview. He urges Gödel to keep quiet but, just in case, also warns Einstein about Gödel's worry before driving them both from Princeton to

---

12 Dyson concludes his comments in the *The Institute Letter*, I.A.S., Spring 2006, p. 6, with: "Now, thirty years later, the observations are far more precise, and we still see no evidence that we live in Gödel's rotating universe."

Trenton. On the way Einstein is telling amusing anecdotes in order to distract Gödel, apparently with great success. The judge at the office, properly impressed by the witnesses, invites all three to attend the (normally private) ceremony. "Up to now," he begins, "you have held German citizenship."- *Austrian*, corrects him Gödel. "Anyhow, it was under an evil dictatorship ... but fortunately, that's not possible in America." *On the contrary*, interjects Gödel, *I know how that can happen*. All three join forces to restrain Gödel from elaborating and to bring the proceedings to their expected conclusion ([F86] p.12, and Zemanek - see [W87] pp. 115-116).

"Einstein was enchanted by Gödel's combination of elegance and precision" ([K80], p. 157). Einstein is such a legend that most people are afraid to approach him. Gödel isn't. There is a feeling of equality between them. Their debates range from the trivial to the profound. Gödel is skeptical about Einstein's idea of a unified field theory and says so ([Y02] p. 56). Straus makes an observation which resonates with our epigraph to Sect. 1: "Gödel had an interesting axiom by which he looked at the world: *nothing that happens is due to accident or stupidity*. If you take that axiom seriously all the strange theories that Gödel believed in become absolutely necessary ... Einstein did not really mind it, in fact thought it quite amusing. Except the last time ... [after the 1952 presidential election] he said: 'You know Gödel has really gone crazy.' So I said, 'Well, what worse could he have done?' 'He voted for Eisenhower.'" (quoted in [W87] p. 32). (Einstein, as most intellectuals, has been for Stevenson.)

One such *strange theory* which illustrates Straus' observation is Gödel's conviction that Leibniz has indeed developed mathematical logic much beyond what we know[13] and his suspicion that some important work of Leibniz has been destroyed by "those who do not want men to become more intelligent." When Gödel's host in Notre Dame, Menger (at whose colloquium in Vienna Gödel has presented 13 contributions) suggests that Voltaire would have been a more likely target, Gödel counters: *Who ever became more intelligent by reading Voltaire?* ([W87] p. 103). Morgenstern is also dubious when Gödel tells him that Leibniz had already discovered the antinomies of set theory "cloaked in the language of concepts but exactly the same" and knew the law of energy conservation, but has been "systematically sabotaged by his editors". Menger and he agree that Gödel is too much alone and that regular teaching duties would be good for him. Then something strange happens, however (which Morgenstern also shares with Menger). Gödel takes him to the Princeton University Library and shows him an abundance of material for comparison: *firstly* books and articles which had appeared during or shortly after Leibniz's life containing exact references to his writings; *secondly* the cited collections. The cited volumes either contain nothing of Leibniz, or the series breaks just before the cited passages, or else the volumes in question have never appeared. Morgenstern has no explanation for these strange facts ([D97] p. 166). Later he tries (without success) to help Gödel obtain copies of some missing references from Europe.

A later example is provided by Gödel's (perfection of Anselm-Descartes-Leibniz) ontological proof of the existence of God. It is based on axioms such as "Being God-like is a positive property" and "Being a positive property is (logical, hence) necessary" ([W87] p. 195[14]). But, Gödel says, *in philosophy he never arrived at what he looked for* ([W87] p. 4). Gödel is confiding his ontological argument to friends during the health (mental?) crisis of 1970, his worst since 1936. He does recover though, and even appears more open[15] (*less formidable*, in the words of an IAS secretary; Kreisel interprets the change as a loss of *his exquisite sense of discretion*, trying to hide his depression, [K80] p. 160). It is during this period, in June 1972, that, attending a conference at IAS dedicated to von Neumann's work on computers, he poses from the floor two questions (the most significant at the meeting, according to Morgenstern): 1. *Is there enough specificity in genetic enzymatic processes to permit a mechanical interpretation of all functions of life and the mind?* (Gödel's opinion on this question is made clear in his letter of March 1974 to the terminally ill Abraham Robinson, the creator of the non-standard analysis: "As you know I have unorthodox view about many things. ... The assertion that our ego consists of protein molecules seems to me one of the most ridiculous ever made..." - cited in [F05] Sect. 5.)

---

13 This is corroborated by Russell who however attributes it to Leibniz's desire "to win the approbation of ... princesses." As a result there is "the popular Leibniz who invented the doctrine that this is the best of all possible worlds (whom Voltaire caricatured as Dr. Pangloss" in *Candide*). "The other, who has been slowly unearthed from his manuscripts by fairly recent editors, was profound, coherent ... and amazingly logical. ... Leibniz's work on mathematical logic would have been enormously important if he had published it" ([R], Book Three, Chapter XI, pp. 581 and 591).

14 See also [SDM] p. 160; on p. 158, there is a photo of Gödel's notebook "Fehler in der Bibel". Kurt writes to his mother (October 1961) "... already today it may be possible purely rationally (without the support of faith ...) to apprehend that the theological world view is thoroughly compatible with all known facts... What I call the theological world-view is the idea that the world and everything in it has a good and indubitable meaning... Since our earthly existence has in itself a very doubtful meaning, it follows directly that it can only be a means toward the goal of another existence. The idea that everything in the world has a meaning is precisely analogous to the principle that everything has a cause on which the whole science rests." - see http://www.edge.org/, The Reality Club, George Dyson, 5.6.06 (letter c/o S. Feferman).

15 But, Morgenstern notes in his diary, whenever one speaks with Gödel one is thrust "immediately into another world".

*2. Is there anything paradoxical in a machine that knows its program completely?* ([D97] p. 243).

After Einstein's death, Gödel is the only Professor at IAS with no car; sometimes he uses the Institute limousine that carries, weekly, members (and family) to the Shopping Centre. His skinny figure at the front seat, wrapped with a shawl, with an overcoat in spite of the warm season, looks strange. (In the spring of 1975 I observed him, while sitting in the same limousine.) It has been noticed that Gödel always feels more comfortable in hot weather. It is, probably, not an accident that he obtains his most famous results during the summers of 1929 and 1930 ([W87] p.99); the crucial step in verifying the continuum hypothesis is done in June 1937 ([W87] p. XXI).

In 1970 Adele's health begins to fail. Kurt no longer may count on her support during his crises of paranoia (he is afraid of getting poisoned and refuses to eat). When Gödel dies (in January 1978) he weighs eighty pounds (some thirty-six kilos – see [SDM] p. 105).

Furtwängler, Gödel's revered professor, is reported to have asked: *Is his illness a consequence of proving the nonprovability or is his illness necessary for such an occupation?* ([Y02] p. 52). This is, perhaps, not just a joke. In the words of his biographer, central to Gödel's life and thought are a few deeply held convictions: (i) *the universe is rationally organised and comprehensible;* (ii) *there is a mental realm apart from the physical world*[16]*;* (iii) *conceptual understanding is to be thought through introspection*; beliefs inspiring both his accomplishments and his angst ([D97] p. 261).

Mathematicians, unlike poets, would more readily accept (ii) than (iii). But how then to perceive new mathematical concepts? Says Gödel: *Positivists contradict themselves when it comes to introspection, which they do not recognize as experience... The concept of set, for instance, is not obtained by abstraction from experience* ([D97] p. 240) – a view resonating with the vision of Cantor, the creator of set theory (Sect. 1). Gödel "conjectures that some physical organ is necessary to make the handling of abstract impressions possible... Such a sensory organ must be closely related to the neural centre for language... For each vague intuitive concept the sharp concept exists all along, only we do not perceive it clearly at first" [W87] p. 190).

*If you understand yourself completely, you understand everything.* One may think that these words refer to Gogol (or to Dostoevsky). Yet, they belong to Gödel ([W87] p. 210).

The saga of over twenty years of labour of interested logicians, mathematicians, historians and philosophers on digging out, deciphering and publishing Gödel's Nachlaß (beginning in 1982 when J. W. Dawson starts cataloguing the ten file cabinets and over fifty cartons in the basement of IAS until the appearance, in 2003, of the fifth, so far last, volume of Gödel's collected work [G] – see [F05]) is not yet finished. There remain over hundred notebooks – in mathematics, logic and foundations, philosophy, theology (including church history) – almost entirely in the obsolete *Gabelsberger* German shorthand. After years of work and some good luck more than half of the 3000 odd pages is transcribed into German from which substantial portions are translated into English. "That's the *good news*. The *bad news* is that what we have as a result is not at all suitable for publication in its present form; after extensive discussion the editors judged it would take a considerable further investment of time, energy and funding to make the material widely available – time, energy and funding that we could no longer draw on either individually or as a group." - writes the editor in chief, [F05]; "No doubt there are many more gems to be unearthed, but we'll have to bequeath them to those with the capacity and inspiration to carry on the work." (More about what remains to be done can be found in [DD05].)

---

16 Answering (in a draft letter) Russel's description of him as an unadulterated Platonist Gödel quotes Russel's own words: "Logic is concerned with the real world just as truly as zoology, though with its most abstract features" ([W87] p. 112).

As Gödel stresses repeatedly our time is not receptive to philosophical thought. *One is stuck by a significant regress in many of the spiritual sciences*, he writes to his mother in 1962 ([W87] p.123). Let us hope that his still unpublished manuscripts will have to wait less than three centuries which elapsed before Leibniz's writings were unearthed.

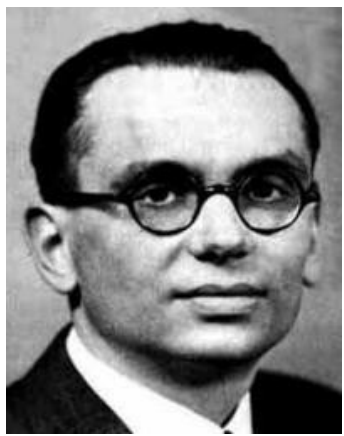

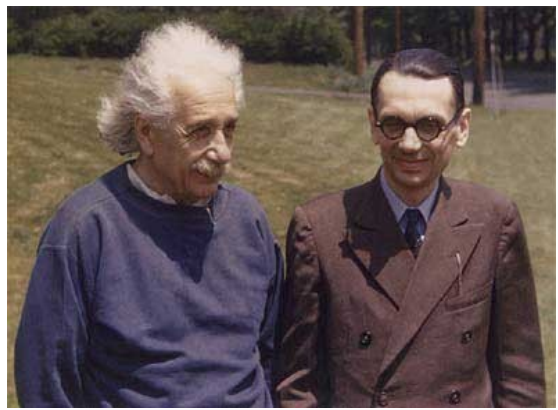

Kurt Gödel and Albert Einstein

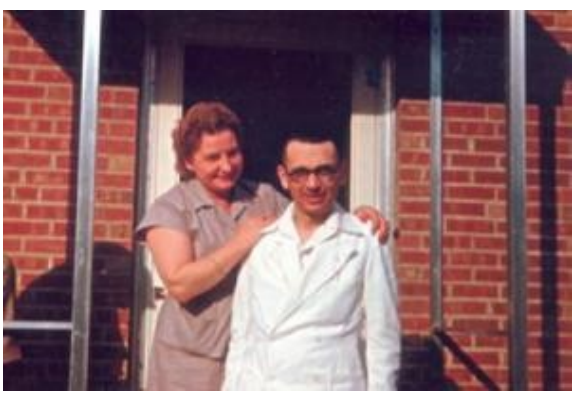

Gödel with wife Adele in Princeton

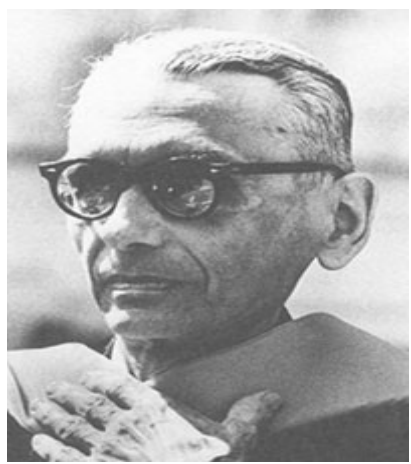